# Resolving structural dynamics *in situ* through cryogenic electron tomography

**Jackson Carrion[1] and Joseph H. Davis[1,2],✉**

1. Computational and Systems Biology Graduate Program
2. Department of Biology

Massachusetts Institute of Technology, Cambridge, Massachusetts, 02139, United States.

✉ Correspondence: jhdavis@mit.edu

**ABSTRACT**

Cryo-electron tomography (cryo-ET) has emerged as a powerful tool for studying the structural heterogeneity of proteins and their complexes, offering insights into macromolecular dynamics directly within cells. Driven by recent computational advances, including powerful machine learning frameworks, researchers can now resolve both discrete structural states and continuous conformational changes from 3D subtomograms and stacks of 2D particle-images acquired across tilt-series. In this review, we survey recent innovations in particle classification and heterogeneous 3D reconstruction methods, focusing specifically on the relative merits of workflows that operate on reconstructed 3D subtomogram volumes compared to those using extracted 2D particle-images. We additionally highlight how these methods have provided specific biological insights into the organization, dynamics, and structural variability of cellular components. Finally, we advocate for the development of benchmarking datasets collected *in vitro* and *in situ* to enable a more objective comparison of existent and emerging methods for particle classification and heterogeneous 3D reconstruction.

---

## INTRODUCTION

Since David DeRosier and colleagues highlighted the promise of combining multiple tilted views to reconstruct a three-dimensional density map of a structure (DeRosier and Klug 1968; DeRosier and Moore 1970), the field of electron tomography has made steady progress toward visualizing macromolecular complexes in three dimensions – first using purified samples and later with vitrified cells. Indeed, early studies, including pioneering work by Baumeister and colleagues, demonstrated the potential of cryogenic electron tomography (cryo-ET) for *in situ* structural analysis by visualizing the coarse-grained architecture of a eukaryotic cell and resolving large, abundant cytosolic complexes, such as the ribosome and proteasome (Medalia et al. 2002). Over the subsequent two decades, cryo-ET has provided insights into cellular organization and architecture from a growing list of systems including prokaryotic, archaeal, and eukaryotic cells, as well as isolated organelles and viruses (Turk and Baumeister 2020; Baumeister 2022; Wang et al. 2023; Waltz et al. 2025). In parallel, ongoing improvements in subtomogram averaging (STA) methods have enabled researchers to resolve key macromolecular complexes *in situ* at resolutions sufficient for the analysis of molecular mechanisms.

Despite these successes, cryo-ET faces inherent challenges that require innovative computational solutions. Compared to single-particle analysis (SPA), the use of tilt-series data in cryo-ET presents both opportunities and challenges for the reconstruction of structurally heterogeneous molecules. In one regard, by providing multiple views of individual particles, tilt-series image acquisition increases projection angle sampling, which is especially beneficial for asymmetric particles that adopt preferred orientations, and it helps distinguish particles positioned along the imaging axis. Simultaneously, it introduces challenges such as reduced signal relative to noise (often described as the signal-to-noise ratio, or SNR) within individual images resulting from the lower electron dose per image, a thicker imaging path at higher tilt angles, and increased radiation damage for images late in the acquisition series. Additionally, tilting the sample introduces defocus gradients across the images that must be taken into account. Finally, when performing tomogram reconstruction, errors in coarse tilt alignment estimation can strongly limit the achievable resolution during tomogram reconstruction and subtomogram averaging (STA) (Bharat and Scheres 2016; Pyle and Zanetti 2021; Zivanov et al. 2022; Tegunov et al. 2021).

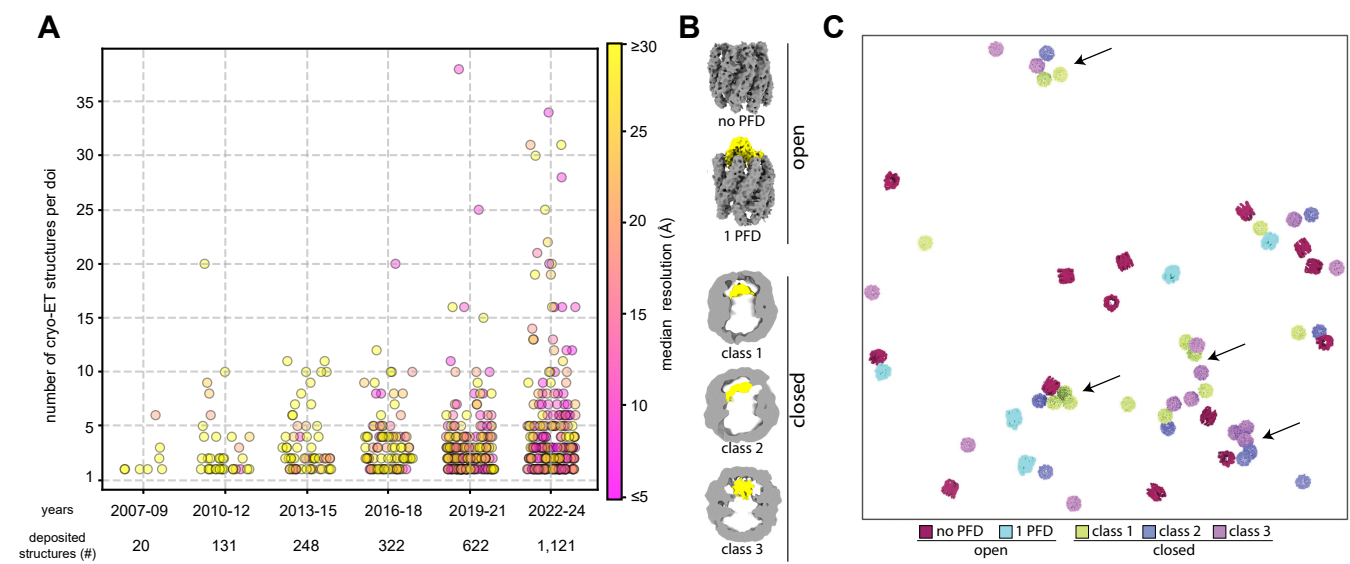

**Figure 1. Capacity and efficacy of *in situ* structural analysis are growing rapidly.**

**(A)** Meta-analysis of Electron Microscopy Data Bank (EMDB) deposited structures resolved by cryo-ET. Each point represents a unique study (DOI) associated with one or more structures produced by subtomogram averaging (STA) that was deposited in the EMDB between 2007 and 2024. The x-axis is binned into three-year intervals, each labeled with the total number of STA structures deposited during that period. The y-axis indicates the number of STA structures deposited per DOI, serving as a proxy for dataset throughput and/or reconstruction diversity within a single study. Points are color-coded by the median resolution of the STA reconstructions associated with each DOI, following color scale (right). Note time-dependent increase in total number of structures, average number of structures per study, and resolution. These data were retrieved from the EMDB in January 2025 and includes all entries annotated under the method "subtomogram averaging."

**(B)** Heterogeneous reconstructions of the TRiC chaperonin complex, highlighting conformational and compositional diversity. The core TRiC structure is colored grey, with variable regions, including a prefoldin domain (PFD) and inferred client proteins, highlighted in yellow.

**(C)** Conformational states depicted in B, mapped to their original positions within the tomogram, revealing the spatial organization and diversity of TRiC processing states in the native cellular environment. Arrows indicate apparent clusters of substrate-loaded open complexes. Panels B-C were adapted from Xing et al. 2025.

Advances in computational methods (Tegunov and Cramer 2019; Tegunov et al. 2021; Zivanov et al. 2022; Liu et al. 2023; Burt et al. 2024) now effectively address many of the inherent challenges in cryo-ET, allowing for the reconstruction of both discrete and continuous structural ensembles of protein complexes spanning scales from the relatively small ~800 kDa ATP-synthase (Dietrich et al. 2024) to multi-megadalton photosynthetic complexes (You et al. 2023). The Electron Microscopy Data Bank (EMDB) is growing exponentially with such work, both in the number of studies published and in the number of structures resolved within a given study, highlighting how researchers are increasingly leveraging the rich data cryo-ET provides (**Figure 1A**). Indeed, through their recent *in situ* analysis of the eukaryotic chaperonin TRiC, which folds nearly 10% of the cellular proteome through dynamic ATP-driven structural transitions, the Beck and Frydman groups showcased this growing capability (Xing et al. 2025). Specifically, they employed a 3D classification method to resolve multiple conformational states of TRiC directly in cells, effectively differentiating TRiC assemblies with one or two prefoldin (PFD) domains observed, and they resolved a conformational change in the chaperonin that closes the chamber, thereby encapsulating its client protein (**Figure 1B-C**).

**Structure determination using 2D particle-image stacks or 3D subtomogram volumes.**

Methods for heterogeneous reconstruction and classification in cryo-ET can be broadly categorized along two key axes (**Table 1**): *1*) those that operate on stacks of 2D particle-images versus 3D subtomogram volumes; and *2*) those aimed at discrete classification versus continuous classification. Whereas the field has traditionally used 3D subtomograms for downstream STA, the increased computational efficiency and the minimization of interpolation errors enabled by the 2D workflows have been more recently adopted. Indeed, early tomographic methods enabled the extraction of 3D sub-volumes from reconstructed tomograms and, as this approach is very intuitive, it found widespread use as practitioners worked to determine coarse molecular structures and to resolve conformational states within those structures using classification-based tools. However, the use of such 3D subtomograms introduces key challenges, including a requirement to interpolate missing data from the limited tilt range – both between tilts and for the entire swath of the 'missing wedge' – and the computational and storage demands of the interpolated three dimensional data (Bartesaghi et al. 2008; Bharat and Scheres 2016; Wan and Briggs 2016; Hagen et al. 2017). The field has recently shifted away from 3D subtomograms and towards using 2D particle-stacks, an approach that reduces storage and computational costs while enabling higher-resolution reconstructions. By operating directly on tilt-series images, tools like Warp and M allow for fine-grained, per-particle refinement of tilt geometry, motion, and CTF parameters during reconstruction—capabilities that are difficult to implement with interpolated 3D subtomograms (Tegunov and Cramer 2019; Tegunov et al. 2021). Tools to resolve structural heterogeneity using such 2D data have also been developed and have proven highly efficacious (Himes and Zhang 2018; Liu et al. 2023; Burt et al. 2024).

**Discrete vs continuous classification.**

The traditional discrete classification methods, such as those implemented in RELION (Zivanov et al. 2022), group particles into a fixed number of predefined classes that, in principle, each represent a specific structural state. This approach is ideal for systems where a limited number of structural states are expected. By concentrating particles into a fixed and relatively small number of classes, classification tends to maximize the number of particles in each class, thereby improving the signal-to-noise ratio of structurally homogeneous regions of the map and ultimately enabling high-resolution reconstructions for the most abundant states. However, forcing particles into discrete classes inherently limits the exploration of structural diversity, and it can result in rare or unexpected states being averaged into broader categories, potentially obscuring biologically significant heterogeneity.

| Tool | Classification | Data Type | Approach to Heterogeneity | Priors | Reference |
|---|---|---|---|---|---|
| RELION 5.0 | Discrete | 2D | Bayesian inference | Pose, CTF, class # | Burt et al. 2024 |
| NextPYP | Discrete | 2D | Linear decomposition | Pose, CTF, class # | Liu et al. 2023 |
| RELION 3.1-4.0 | Discrete | 3D | Bayesian inference | Pose, CTF, class # | Bharat and Scheres 2016 |
| STOPGAP | Discrete | 3D | Linear decomposition | Pose, CTF, class # | Wan et al. 2024 |
| tomoDRGN | Continuous | 2D | Machine learning | Pose, CTF | Powell and Davis 2024 |
| CryoDRGN-ET | Continuous | 2D | Machine learning | Pose, CTF | Ragnan et al. 2024 |
| emClarity | Continuous | 2D | Linear decomposition | Pose, CTF | Himes and Zhang 2018 |
| OPUS-TOMO | Continuous | 3D | Machine learning | Pose, CTF | Luo et al. 2024 |
| Dynamo | Continuous | 3D | Linear decomposition | Pose, CTF | Castaño-Díez et al. 2012 |
| PEET | Continuous | 3D | Linear decomposition | Pose, CTF | Heumann et al. 2011 |
| MDTOMO | Continuous | 3D | Normal mode analysis of MD simulation | Pose, CTF, atomic model | Vuillemot et al. 2023 |
| TomoFlow | Continuous | 3D | Dense optical flow | Pose, CTF | Harastani et al. 2022 |

**Table 1. Overview of particle classification and heterogenous reconstruction tools.**
Categorization of methods for cryo-ET heterogeneity analysis based on their key attributes, including: whether they aim for discrete or continuous classification; whether they operate on stacks of 2D particle-images or 3D subtomograms; the underlying approaches to heterogeneity; and the additional prior information required.

In contrast, continuous classification methods map particles into a low-dimensional 'latent' space, or manifold, which allows for the representation of a conformational continuum. This approach is particularly advantageous when analyzing systems with complex and variable structural landscapes bearing an unknown, or unknowable, number of structural states. Indeed, by forgoing the need to restrict particles to discrete categories, continuous classification has the potential to capture a broader range of conformations, and to provide a more comprehensive view of the structural landscape *in vitro* and *in situ* (Vuillemot et al. 2023; Powell and Davis 2024; Rangan et al. 2024). It is notable however, that methods for continuous classification typically use a vastly increased parameter space, and thus these methods must balance attempts to represent a full structural continuum against overfitting to the high levels of noise inherent in cryo-ET. Conversely, when using too few classes, underfitting can occur where class-based averaging can obscure important structural differences. This can, in turn, lead to an oversimplified representation of molecular motion and compositional variability, limiting insights into biologically relevant conformational states.

**Methods operating on 3D subtomograms vs 2D particle stacks.**
The introduction of STA workflows in RELION 3.1 (Bharat and Scheres 2016) marked a significant milestone for the field of cryo-ET, as it extended the single-particle analysis (SPA) workflow, which was familiar to many researchers, to subtomogram volumes. Specifically, individual 3D subtomograms extracted from a tomogram could be sorted amongst a fixed number (commonly ~3-6) of discrete classes. Once assigned, individual classes could be further refined often resulting in higher resolution reconstructions. Additionally, integration of such subtomograms into standard RELION workflows allowed for the application of real space masks, allowing users to focus the classification on specific regions of interest, which has proven valuable in identifying distinct structural states across an array of macromolecules (Li et al. 2019; Xing et al. 2023; Woldeyes et al. 2023; You et al. 2023; Basiashvili et al. 2023; Ruehle et al. 2024; Li et al. 2024; Khusainov et al. 2024; Klumpe et al. 2024; Singh et al. 2024; Watanabe et al. 2024). Notably, this classification and refinement strategy can be performed iteratively and hierarchically, allowing one to better reveal subtle but important structural differences (Gemmer et al. 2023; Fedry et al. 2024; Gemmer et al. 2024; Pyle et al. 2024; Pražák et al. 2024; Wang et al. 2024) using a relatively small number of classes in each round of classification, which is vital both for computational efficiency and to increase the effective number of particles assigned to each class.

Relatedly, STOPGAP (Wan et al. 2024) performs STA and discrete classification within the 3D subtomogram space. It employs a stochastic hill-climbing algorithm complemented by simulated annealing (Reboul et al. 2016) during early alignment stages that, in combination, aim to avoid local minima during orientation and class assignment. Classification typically begins with a set of initial reference volumes, either provided manually or generated by randomly partitioning particles into subsets. These references serve as the starting point for iterative classification and refinement, where each round of classification involves aligning subtomograms to current references. Due to the stochastic nature of these methods, the classification can be run repeatedly, and STOPGAP can identify and retain particles that consistently classify into the same group while excluding inconsistently assigned particles. This process, known as consistency cleaning, ensures that only the most reliable particles are included in the final classes. Combined, it is thought that these features make STOPGAP particularly effective at classification, and it has been widely used (Khavnekar et al. 2023; Wagner et al. 2024; Taniguchi et al. 2024; Datler et al. 2024).

More recently, the aforementioned advantages of working directly with 2D tilt-series images have become more widely appreciated, and newly developed methods have increasingly adopted this approach. RELION 5.0 (Burt et al. 2024) allows for a full 2D particle-stack workflow through its "*extract subtomos*" job that combines particle positions and tomogram alignment parameters to crop regions around particles in the tilt series micrographs, thereby generating stacks of 2D particle-images. To address challenges related to the wide range of image quality often observed across a stack of 2D tilt-images, RELION 5.0 includes options to exclude images above a user-defined electron dose threshold and to filter particles that are not visible in a minimum number of tilt images. These features help reduce the impact of variable electron dose and missing views. Once processed, these stacks can be used in the traditional RELION pipeline, including *ab initio* reconstructions, refinement, discrete 3D classification, and others (Burt et al. 2024; Gonzalez-Magaldi et al. 2024; Kelley et al. 2024; Isbilir et al. 2024).

NextPYP (Liu et al. 2023) further builds on the 2D particle-image concept, offering a scalable, end-to-end framework for cryo-ET data analysis that eliminates the need to generate large tomograms or subtomogram-volumes. Instead, NextPYP processes raw, unaligned tilt series images directly, extracting particle images on-the-fly as needed and thus bypassing the need for intermediate particle stacks. This on-the-fly extraction approach further reduces the storage footprint of cryo-ET datasets and NextPYP's seamless integration of tools for motion correction, CTF estimation, particle picking, 3D reconstruction, and analysis of structural heterogeneity through discrete and continuous (see below) methods significantly accelerates data processing.

**Continuous classification: linear decomposition.**
Whereas discrete classification methods have been foundational in the development of STA, they are inherently misaligned with macromolecular complexes that undergo continuous conformational changes, or those that adopt a large number of discrete states. For researchers aiming to resolve continuous conformational transitions or explore more complex structural landscapes, continuous classification and reconstruction strategies are essential (Kinman et al. 2023; Sun et al. 2023). These approaches can be broadly categorized into traditional linear decomposition methods and more advanced non-linear algorithms, including machine learning-based methods, with variants of each developed to operate on 3D subtomograms or on stacks of 2D particle-images.

Principal component analysis (PCA) and singular value decomposition (SVD) are foundational linear decomposition techniques that have been applied in cryo-ET suites including Dynamo (Castaño-Díez et al. 2012), i3 (Winkler 2007), and PEET (Heumann et al. 2011), which have historically supported PCA-based heterogeneity analysis. These techniques decompose sets of extracted 3D volumes into orthogonal components, enabling the identification of dominant modes of variability. By analyzing these modes, researchers can map particles onto a continuous conformational space, providing a low-dimensional representation of heterogeneity. Many of the software packages using these methods interlace reconstructions and classifications by calculating the covariance matrix of aligned particles during each iteration. PCA on this matrix projects the data into a low-dimensional space defined by the dominant modes of structural variance. This projection facilitates the visualization and exploration of particle distributions, potentially highlighting groupings related to underlying structural differences, which can be clustered or sampled dynamically (Zhang 2019; Pyle and Zanetti 2021; Kaplan et al. 2023; Richard et al. 2024; Sibert et al. 2024). Additionally, Dynamo and PEET allow users to specify spatial masks, enabling the classification of specific regions that exhibit structural variability.

Whereas most approaches rely on subtomogram averaging and iterative classification to boost the signal-to-noise ratio, methods like MDTOMO (Vuillemot et al. 2023) and HEMNMA-3D (Harastani et al. 2021) adopt a different strategy by operating directly on individual subtomograms. To mitigate the impact of the inherently higher noise for single subtomograms, these methods integrate physics-based priors, such as normal mode analysis (NMA) or molecular dynamics (MD) simulations, to guide flexible fitting of atomic models into subtomograms. This enables the recovery of continuous heterogeneity at the atomic level, constrained by energetically and structurally plausible motions. The resulting deformation parameters are then projected into a low-dimensional space using PCA, as described above. This framework enables researchers to recover prominent structural transitions in terms of modeled atomic displacements, and to cluster or average density maps at specific positions along the energy landscape, offering a detailed and interpretable view of continuous heterogeneity.

Complementing the linear decomposition approaches applied directly to ensembles of 3D subtomograms, the emClarity approach developed by the Zhang group (Himes and Zhang 2018) incorporates biologically meaningful priors into the analysis, but in the 2D tilt-series space, rather than in the 3D subtomogram space. By applying band-pass filters, emClarity computes intervoxel correlations at biologically relevant length scales, targeting structural features such as ~10 Å for α-helices, ~20 Å for RNA helices, and ~40 Å for protein domains. SVD is then applied independently at each scale, and the resulting dominant components are concatenated into a feature vector for downstream clustering and classification. Additionally, emClarity incorporates particle-based tilt-series geometry refinement, similar to the refinement methods described in Warp (Tegunov and Cramer 2019) and M (Tegunov et al. 2021). This iterative process refines both particle alignments and the geometric parameters of the tilt series using local, patch-based optimization, allowing the correction of regional tilt deformations while using regularization to limit overfitting. These refinements are most naturally applied to the 2D tilt-images, highlighting a key advantage of 2D image-based methods.

**Continuous classification: machine learning-based approaches.**
Machine learning architectures, particularly autoencoders (Kingma and Welling 2013), have emerged as powerful tools for analyzing cryo-ET data (Schwalbe et al. 2024). In a typical autoencoder architecture, input data (*i.e.* 2D tilt-series images or 3D subtomograms) are passed through an encoder network, which compresses the high-dimensional structural information into a simplified, low-dimensional latent space. This latent representation captures variations across particles, both compositional and conformational and, ideally, discards noise. A decoder network then reconstructs the input from this latent representation. Critically, this model can be trained in a self-supervised manner by comparing the decoder's output to the input. Once trained, the autoencoder can generate particle-specific reconstructions. Notably, because each particle is mapped to a distinct point in the continuous latent space, these models avoid the rigid boundaries of discrete classification and are particularly effective in resolving continuous conformational changes and in identifying sparsely populated states.

Several recent tools adopt this autoencoder framework, though they differ in their architectural design and data handling strategies. For example, tomoDRGN (Powell and Davis 2024) and cryoDRGN-ET (Rangan et al. 2024) extend the original single-particle analysis (SPA) cryoDRGN (Zhong et al. 2019, 2021; Kinman et al. 2023) variational autoencoder framework to operate on stacks of 2D tilt-series images. Note that using the 2D tilt-series data introduces specific challenges not present in SPA workflows, as each particle is associated with multiple tilt images, each with a distinct CTF, cumulative electron dose, and SNR characteristics. Additionally, some tilt images may be missing or unusable for a substantial fraction of particles analyzed due to issues associated with occlusion of the particle by another molecule along the imaging axis, poor contrast, or technical failures in image acquisition or alignment. To address this, tomoDRGN randomly samples a subset of available tilt images for each particle during training. This enforces uniform input dimensionality to the model during training while flexibly handling missing tilt-images and leveraging all available data across a tilt-series. In contrast, cryoDRGN-ET, uses a relatively small, fixed set of high-quality images - typically low-tilt angles - and discards any particles that lack this defined set of views. OPUS-TOMO (Luo et al. 2024) takes yet another approach by operating on real space 3D subtomograms using a convolutional encoder-decoder architecture.

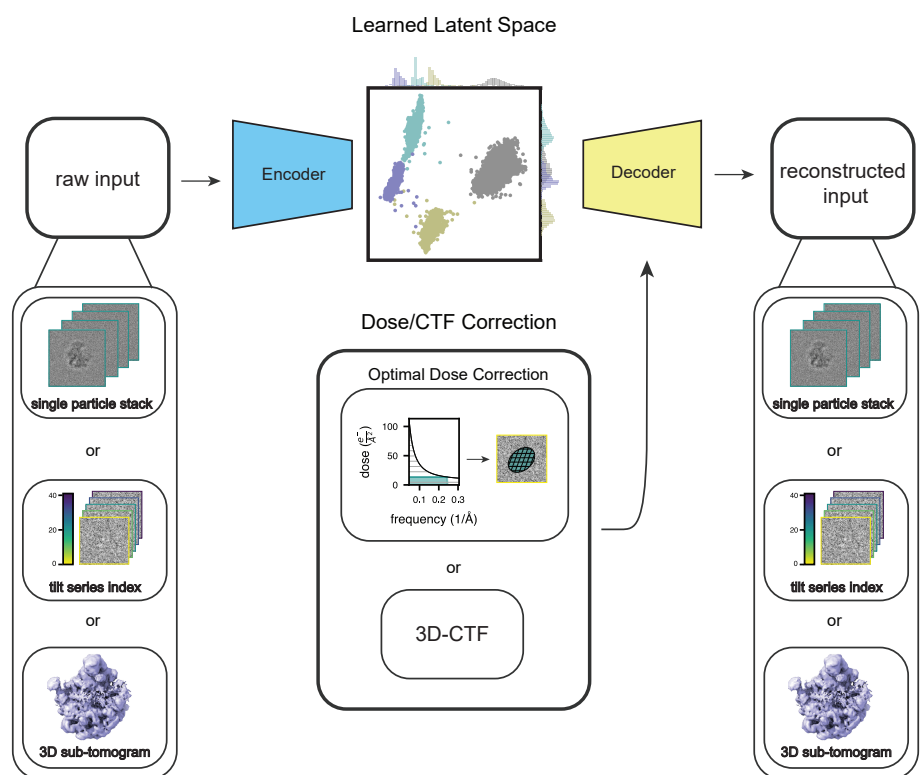

**Figure 2. Autoencoder frameworks applied to cryo-ET datasets.**
A schematic illustration of the autoencoder framework applied to cryo-ET data, highlighting its versatility across different input types (left). The encoder (blue) compresses each input into a learned low-dimensional latent space, ideally sorting individual particles based on their structural differences. Each point in the latent space corresponds to a single particle, which can reveal distinct clusters or continuous conformational landscapes. The decoder (yellow) reconstructs the input from the corresponding point in latent space, allowing for self-supervision during network training. Once trained, researchers can readily sample 3D structures from various regions of latent space, enabling the reconstruction of a continuous distribution of density maps. Adapted from (Powell and Davis 2024).

To manage variation in image quality across the tilt series, tomoDRGN weights images by tilt angle to emphasize low-tilt, high-SNR views. Because the model operates in reciprocal space, it also applies frequency-dependent masking to exclude high-frequency components from tilt-images degraded by high cumulative electron dose. Together, these strategies improve both reconstruction fidelity and computational efficiency (Powell and Davis 2024; Powell et al. 2025).

**What lies ahead: the need for benchmarking datasets.**
As seen in the adjacent fields of image processing (Lecun et al. 1998) and protein structure prediction (Moult et al. 1995), the creation of robust benchmarking datasets will be critical to assess and improve tools for analyzing structural heterogeneity in cryo-ET. Indeed, during the development of heterogeneous structural analysis methods in SPA, public datasets (Davis et al. 2016; Plaschka et al. 2017) with varying degrees of structural heterogeneity served (and continue to serve) as standard benchmarks for evaluating new tools. In both instances, these datasets were especially valuable because experts deeply familiar with the systems had curated and annotated particle populations based on complementary biochemical data, providing a functional '*ground truth*' against which newly developed tools could be easily compared. Simulated datasets (Fu et al. 2007; Lyumkis et al. 2013) have long complemented these experimentally derived (*i.e.* '*real*') ones as they naturally afford per-particle ground truth structures for assessing reconstruction accuracy, although they often fail to faithfully simulate the noise and imaging artifacts common in real data. Specific metrics, including reconstruction resolution and classification accuracy, evaluated on a per-particle basis, when ground truth is known, or in terms of the correct population frequency, have enabled comparison and iterative improvement of a large swath of heterogeneous single-particle analysis methods. Recently, such simulated and real resources were consolidated in a "cryoBench" (Jeon et al. 2024).

A similar paradigm is emerging in cryo-ET, where both simulated tilt-series datasets (Powell and Davis 2024; Kinman et al. 2025a) and increasingly well-annotated real tilt-series datasets from purified particles (Danev et al. 2014; Schur et al. 2016; Bharat and Scheres 2016) and cells (Tegunov et al. 2021) are often analyzed. Notably, a recent large-scale cryo-ET study of Chlamydomonas reinhardtii cells using plasma-based focused ion beam milling produced ~1,800 tomograms, and the accompanying manuscript details expert-guided analysis of both soluble and membrane-bound complexes ranging from ~200 kDa to over 3 MDa (Kelley et al. 2024). As this dataset includes a variety of curated macromolecular assemblies, many of which exhibit dynamic conformational states, it is likely to be a particularly rich resource for benchmarking tomographic reconstruction and heterogeneity analysis methods.

Our group recently developed SPA benchmarking datasets that merge real and synthetic approaches: a related series of multimeric complexes differing by a single 'programmable' component were imaged separately and their extracted particle images were merged into unified dataset (Kinman et al. 2025b). Crucially, by construction, in these datasets the identity of each particle's variable component is known, thus providing ground truth at the level of individual particles, while still retaining the realistic noise and imaging artifacts inherent in real data. Acquiring tilt-series data of this type should be feasible, and future efforts to generate analogous datasets in cellular cryo-ET could serve as definitive benchmarks for evaluating tools aimed at resolving structural heterogeneity *in situ*. Ultimately, the continued development of such benchmarking data will be essential in motivating and evaluating forthcoming methodological innovations aimed at resolving structural dynamics *in situ* through cryogenic electron tomography.

## ACKNOWLEDGEMENTS

We thank Maria Carreira and other members of the Davis lab for helpful discussion and feedback. This work was supported by NIH grant R01-GM144542, NSF-CAREER grant 2046778, the Sloan Foundation, the Smith Family Foundation, and the Whitehead Family.